\documentclass[aoas,MSNbibl,nameyear,dvips]{arximspdf}
\usepackage{dcolumn}
\usepackage{graphicx}

\doi{10.1214/12-AOAS553} %kopijuoti is PTS
\volume{6}
\issue{3}
\pubyear{2012}
\firstpage{994}
\lastpage{1020}

\makeatletter

\newcolumntype{d}[1]{D{.}{.}{#1}}

\newcommand{\xrightarrow}[1]{\stackrel{#1}{\rightarrow}}

\makeatother

\begin{document}
\begin{frontmatter}

\title{Fibre-generated point processes and fields of~orientations}
\runtitle{Fibers, point processes, fields of orientations}

\begin{aug}
\author[A]{\fnms{Bryony J.} \snm{Hill}\ead[label=e1]{Bryony.Hill@gmail.com}},
\author[A]{\fnms{Wilfrid S.} \snm{Kendall}\ead[label=e2]{W.S.Kendall@warwick.ac.uk}}
\and
\author[A]{\fnms{Elke} \snm{Th{\"o}nnes}\corref{}\ead[label=e3]{E.Thonnes@warwick.ac.uk}}
\runauthor{B. J. Hill, W. S. Kendall and E. Th{\"o}nnes}
\affiliation{University of Warwick}
\address[A]{Department of Statistics\\
University of Warwick\\
Coventry, CV4 7AL\\
United Kingdom \\
\printead{e1}\\
\hphantom{E-mail: }\printead*{e2}\\
\hphantom{E-mail: }\printead*{e3}} %adresu isvedimo komanda gale!
\end{aug}

% HISTORY:
\received{\smonth{9} \syear{2011}}
\revised{\smonth{2} \syear{2012}}

% ABSTRACT
%
\begin{abstract}
This paper introduces a new approach to analyzing spatial point data
clustered along or around a system of curves or ``fibres.'' Such data
arise in catalogues of galaxy locations, recorded locations of
earthquakes, aerial images of minefields and pore patterns on
fingerprints. Finding the underlying curvilinear structure of these
point-pattern data sets may not only facilitate a better understanding
of how they arise but also aid reconstruction of missing data. We base
the space of fibres on the set of integral lines of an orientation
field. Using an empirical Bayes approach, we estimate the field of
orientations from anisotropic features of the data. We then sample from
the posterior distribution of fibres, exploring models with different
numbers of clusters, fitting fibres to the clusters as we proceed. The
Bayesian approach permits inference on various properties of the
clusters and associated fibres, and the results perform well on a
number of very different curvilinear structures.
\end{abstract}

% KEYWORDS
%

\end{frontmatter}

%s1 #&#
\section{Introduction}

In this paper we introduce a new empirical Bayes approach concerning
point processes that are clustered along curves or ``fibres,'' with
additional background noise.

\begin{figure}
\begin{tabular}{@{}c@{\quad}c@{}}

\includegraphics{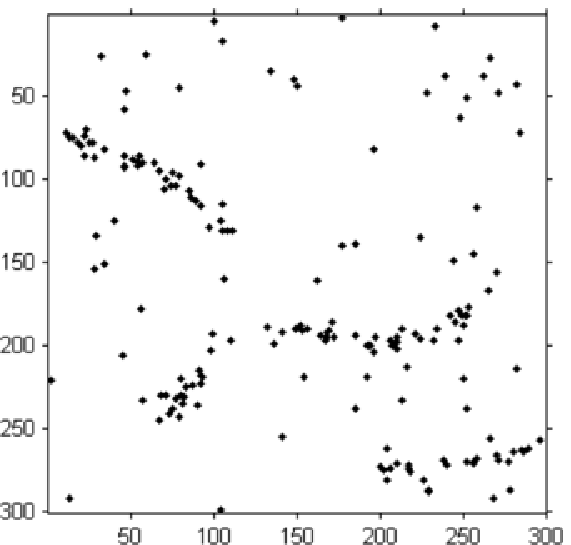}
 & \includegraphics{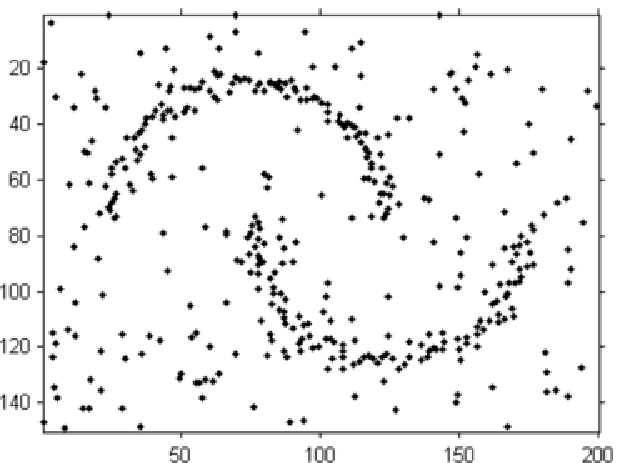}\\
(a) & (b)\\[4pt]

\includegraphics{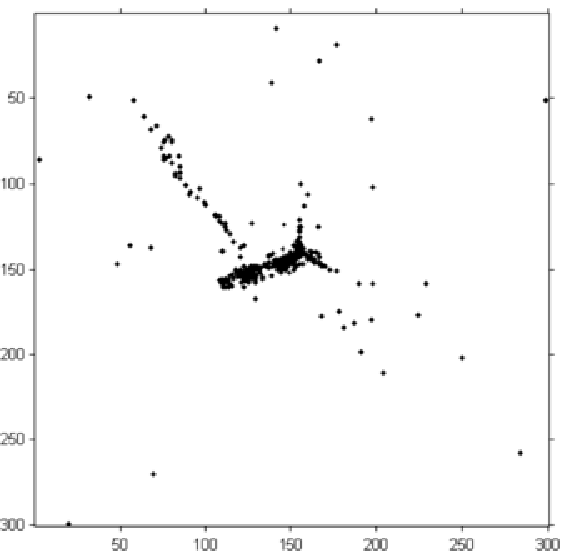}
 & \includegraphics{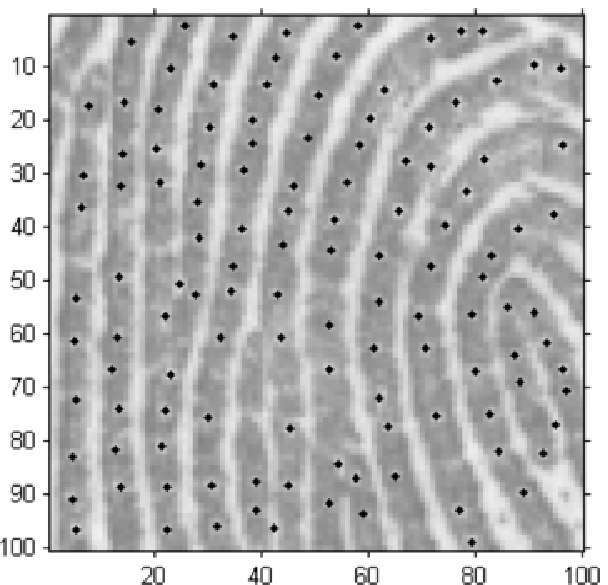}\\
(c) & (d)
\end{tabular}
\caption{Four examples of point patterns clustered around latent
curvilinear features with background noise.
\textup{(a)} Simulated point pattern. \textup{(b)} Simulated point
pattern described in Stanford and Raftery (\protect\citeyear{Sta00}).
\textup{(c)} Earthquake epicenters in the New Madrid region. Data is
taken from CERI (Center for Earthquake Research and
Information). \textup{(d)} Pores along ridges of a portion of the
fingerprint a002--05 from the \textup{NIST Special Database 30}
[Watson (\protect\citeyear{Nist})].}
\label{figExample}
\end{figure}

In nature such point patterns often arise when events occur near some
latent curvilinear generating feature.
For example, earthquakes arise around seismic faults which lie on the
boundaries of tectonic plates and hence are naturally curvilinear.
Similarly, sweat pores in fingerprints lie on the ridges of the finger,
which possess a curvilinear structure.
% Examples of
Figure~\ref{figExample} presents
these data
together with
% and 2
two simulated examples of point patterns clustered around underlying
families of curves with additional background noise.
% are given in Figure~\ref{figExample}.
% Identifying these
Identification of
curvilinear elements and
% elucidating
elucidation of
their relationship with the point data is both an interesting
theoretical problem and a useful tool for gaining understanding of the
origins of the data.
It also provides a technique for reconstruction of missing
data.

%f1 #&#
%

The model introduced here describes families of nonintersecting curves
via
a field of orientations (a map $\upsilon_{\mathrm{FO}}\dvtx W\rightarrow
[0,\pi)$
assigning an undirected orientation to each point in the window).
% This is achieved by considering
The curves are produced as
segments of streamlines integrating the field of orientations.
We say that a curve \textit{integrates} the field of orientations if
the curve is continuously differentiable and of unit speed, and
if\vadjust{\goodbreak}
its tangent agrees with the field of orientations at each point.
The term \textit{streamline} is used to describe a curve which
integrates the field of orientations and has no end points in the
interior of the window $W\setminus\partial{W}$.

We
choose to
use a variant on an empirical Bayes approach to estimate the field of
orientations,
since
% as
a fully Bayesian approach would involve infinite-dimensional
distributions and be very computationally intensive.
% Instead we estimate
The empirical Bayes component consists of estimation of
the field of orientations from the data via a tensor field as
detailed in Section~\ref{secgradfieldall}.
In the following, a tensor field is represented
% Represented
by
% an
assignation
% to each point
of a symmetric nonnegative definite matrix
to each point of the planar window.
%, tensor
Tensor fields of this kind play an important role in diffusion tensor
imaging [DTI, see \citet{Bih01}]. The field of orientations is
constructed simply by calculating the orientations of the
representative matrices' principal eigenvectors; singularities in the
field of orientations correspond to points where there is equality of
the two eigenvectors.

We show how properties of the underlying distribution of fibres can be
estimated using Monte Carlo techniques applied to the spatial point data.
Our approach has the advantage that it can be used to quantify
uncertainty on a range of parameters and does so effectively for
different types of curvilinear structure.
The use of a field of orientations to identify fibres leads to a~strong
performance on data such as that shown in Figure~\ref{figExample}(d),
where there is noticeable alignment of points perpendicular to the fibres.
% is noticeable.

%s1.1 #&#
\subsection{Potential applications}

Point patterns with a latent curvilinear structure
arise
% exist
in many different areas of study.

In seismology, epicenters of earthquakes are typically densely
clustered around seismic faults.
The earthquake data from the New Madrid region as shown in Figure
\ref{figExample}(c) consists of one short dense cluster of points, one
longer rather sparse cluster, both connected, and a relatively small
number of ``noise'' points scattered over the window.
The New Madrid earthquake data is considered in
Stanford and Raftery's (\citeyear{Sta00}) approach
to detecting curvilinear features.

In cosmology, galaxies appear to cluster along inter-connected
filaments forming a three-dimensional web-like structure with large voids
between the filaments.
There is interest in identifying the nature of the filaments [see,
e.g., \citet{Sto07}].
There is also evidence that these galaxies form surfaces or ``walls'' in
some regions. This
suggests
% may lead to
the exciting challenge of
%potentially
extending our model to include two-dimensional surfaces in
three-dimensional space.

A further application is that of sweat pore patterns on fingerprint
ridges [see Figure~\ref{figExample}(d)]. Sweat pores are tiny apertures
along the ridges where the ducts of the sweat glands open.
Robust inference of the ridge structure from the pore pattern has\vadjust{\goodbreak}
potential for aiding reconstruction of patchy fingerprints and may also
allow for more efficient storage of fingerprints in very large databases.
The underlying curve structure is a dense set of locally parallel
curves along which pores are located, usually very close to the center
of the ridges. The noise arises mostly from artifacts in the automatic
extraction of pores from the image.

An issue with fingerprint pore data is that pores usually align across
the ridges as well as along them.
This can complicate the reconstruction of ridges, as the dominant
orientation is less clear.
We overcome this issue by constructing a smooth tensor field which
extrapolates dominant orientation estimates over the regions of
directional ambiguity.

%s1.2 #&#
\subsection{Background}
An existing method of estimating the curves in the underlying structure
of a point process is
Stanford and Raftery's (\citeyear{Sta00}) use of principal curves (a nonlinear generalization of
the first principal component line).
An EM-algorithm is used
%to optimize the model for a number of
%combinations of smoothness and a number of components.
to optimize the model over a variety of choices of smoothness
parameter and number of components.
An optimal choice of smoothness and number of components is then
selected using Bayes factors.
This technique generally performs very well; however, it is sensitive
to the initial clustering of the data
and therefore has difficulties reconstructing fibres in some regions
where fibres may be expected
but signal points are absent [e.g., the fingerprint pore
data---Figure~\ref{figExample}(d)].

A piecewise linear ``Candy model'' (or ``Bisous model'' in three
dimensions) is used by \citet{Sto07} to model filaments in galaxy data.
They compare the empirical densities of galaxies within concentric cylinders
and thus delineate these filaments. This approach is restricted to
piecewise linear fibre models where the deviation of points from fibres
follows a uniform distribution over a thin cylinder centered along the fibre.
Sufficient statistics of the model for data with filamentary
structure are then compared to sufficient statistics on structureless
data sets; see \citet{Stoica-2005} and
Stoica, Mart{\'{i}}nez and Saar
(\citeyear{Sto07,Stoica-2010}).

Density estimates of the point pattern can be obtained using techniques
such as kernel smoothing.
Fibres can be directly estimated from this density; an example of this
can be seen in \citet{Gen09} where steepest ascent paths along the
density estimate are constructed and the density of these paths is analyzed.

A further approach discussed in \citet{Bar85} is based on construction
of the minimal spanning tree of the set of points. In three dimensions
this gives a useful insight into the overall characteristics of the
filamentary structure.

The method presented in \citet{August-2003} is based on a random curve
model in which curvature
is defined as a Brownian motion. The resulting model is used to
enhance contours in the output of edge operators applied to digital
images and thus to data in which signal points
are dense along curvilinear structures.\vadjust{\goodbreak}

The treatment advocated here is also based on the formulation of a
general model for families of curves and the point patterns clustered
around them. In contrast to \citet{August-2003}, curves are
modeled as segments of streamlines integrating a smooth field of
orientations which encourages interpolations over areas of missing
data. The prior model for the orientation field is derived via an
empirical Bayes step. Then birth--death MCMC is used to sample from the
posterior distribution of the fibres. The model formulation itself uses
the initial exploratory work of \citet{Tho08} [see also
\citet{Su09}], which focused on the fingerprint pore data and
described the use of tensor fields for estimating dominant orientations
in spatial point data.\looseness=-1

%s1.3 #&#
\subsection{Problem definition}

In particular, we are interested in modeling a~random point process
$\Pi $ viewed in a planar window $W\subset\mathbb{R}^2$; we write the
observed part of the point process as $W\cap\Pi= \{y_1,\ldots,y_m\}$
for some arbitrary ordering of points. The point process arises from a
mixture of homogeneous background noise and an unknown number of point
clusters, each clustered along a curve, henceforth called a fibre. Thus
a fibre is a one-dimensional object, a smooth curved segment, embedded
in a higher-dimensional space (the space containing the point process).
Random sets of fibres or ``fibre processes'' are discussed in
\citet{Sto95} and \citet{Ill08}.

Having specified an appropriate model, we must identify a suitable
method of analysis of the posterior distribution of fibres given a data
set of spatial point locations $\{y_1,\ldots,y_m\}$ over the window $W$.

%s1.4 #&#
\subsection{Plan of paper}

The paper is laid out as follows.
The following section gives an overview of the model proposed in this paper.
% Full details including
Details of
the
underlying
probability
model
% distributions
are given in Section~\ref{secprobmodel}.
% An outline of the
The
empirical Bayes method of estimating an appropriate field of
orientations is
outlined
% given
in Section~\ref{secgradfield}.
Section~\ref{secBDMCMC} presents a method of sampling from the
posterior distribution of the fibres given the point pattern data using
Monte Carlo methods.
This is implemented for a number of examples in Section~\ref{secexamples}.
In the final section we
compare
% discuss the advantages of
this model
to
% over
other approaches,
discuss
some known issues of implementation and statistical analysis, and
note
possible directions in which this model might be extended.

%s2 #&#
\section{Basic considerations}

We use a Bayesian hierarchical model to describe the relationship
between the points and the fibres.

%f2 #&#
%
\begin{figure}

\includegraphics{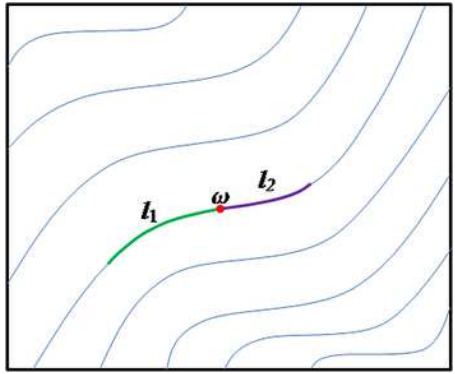}

\caption{An orientation field is depicted as thin grey lines. A fibre
$F(\omega, l_{1}, l_{2})$ is defined as the curve segment that
integrates the orientation field
from reference point $\omega\in W$ in one direction to a distance
$l_{1}$ and in the other direction to a distance
$l_{2}$. Recall, a curve segment is said to integrate the orientation
field if at any point of the segment its tangent
agrees with the orientation field.}\vspace*{-3pt}
\label{figFibreParameters}
\end{figure}

%s2.1 #&#
\subsection{Points}
A natural choice is to model the spatial point process as a~mixed
Poisson process or ``Cox process'' driven by a random fibre process.
By this we mean that the points arise independently and are associated
in some way with a random fibre---typically clustered around it.
Such a point process is called a ``fibre-process generated Cox
process;'' see \citet{Ill08}.\vadjust{\goodbreak}

In our model we do associate points with particular fibres but we
remove the
Poissonian character of the distribution of points along fibres,
replacing this by a renewal process based on Gamma distributions for
interpoint distances.
This allows us to model a tendency to regularity in the way in which
points are distributed
along a fibre but includes the Poissonian case with exponentially
distributed interpoint
distances.\vspace*{-3pt}

%s2.2 #&#
\subsection{Fibres}
\label{secmodelfibres}
In contrast to previous work, in which curves are often constructed as
splines fitted to the data, we define them as integral curves of
a~field of orientations.
This means that at any point on a fibre, the tangent to the fibre
agrees with the field of orientations at that point.
Note that a field of orientations is equivalent to a vector field
except that each point in the field is assigned a
\textit{directionless} orientation.
An instance of a random field of orientations $\Upsilon_{\mathrm{FO}}$ is
written as $\upsilon_{\mathrm{FO}}\dvtx W\rightarrow[0,\pi)$, where
$[0,\pi
)$ represents the space of planar directions (with $0$ and $\pi$ identified).

The simplest way to determine a fibre $F$ is to choose a reference
point $\omega\in W$ on the fibre and
specify the two arc lengths $l_1, l_2 \in{\mathbb R^+}$ of $F\setminus
\{\omega\}$; see Figure~\ref{figFibreParameters}.
For a fixed field of orientations this will characterize a fibre,
although the parametrization by reference point and length is evidently
not unique.
We model the fibres in terms of these parameters (the reference points,
arc lengths and field of orientations).
Note that an alternative construction can be based on random selection
of a finite number of fibres generated by decomposing the streamlines
according to Poisson point processes distributed along the streamlines;
however, this
construction introduces intriguing measure-theoretic issues which are
out of place in the present treatment.

We note that taking the reference points to be uniformly distributed
over the window $W$ will lead to a bias\vadjust{\goodbreak} in the distribution of fibres
in that the mean length of fibre per unit area is not constant across $W$.
This issue has been considered and a solution involving an adjustment
to the distribution of reference points has been identified. We have
not applied the bias correction to our examples, as there is sufficient
data to make the bias negligible.

The field of orientations is a useful intermediary in constructing
fibres and, as such,
is part of a useful decomposition of the construction problem.
In practice, we seek to identify a suitable field of orientations by
analysis of properties of the data.

%s2.3 #&#
\subsection{Noise}
Finally, we include background noise in the form of an independent
homogeneous Poisson process superimposed onto the fibre-generated
``signal'' point process.

%s3 #&#
\section{Probability model}
\label{secprobmodel} 
\begin{figure}

\includegraphics{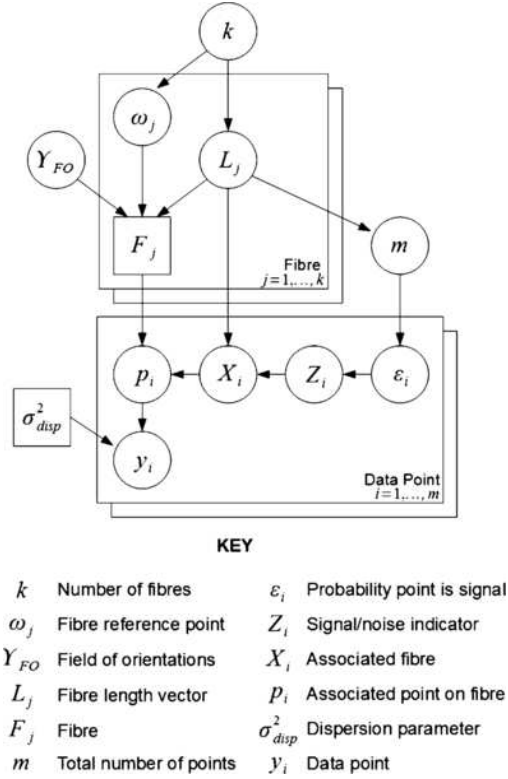}

\caption{Directed acyclic graph (DAG) of model: arrows indicate
conditional dependencies, elements in squares are deterministically
calculated or constant, while those in circles are random variables.
For simplicity we have not included reference to hyperparameters
$\lambda$, $\kappa$, $\eta$, $\alpha_{\mathrm{signal}}$ and
$\beta_{\mathrm{signal}}$.} \label{figDAG}
\end{figure}

A directed acyclic graph (or DAG) showing the
conditional dependencies for the model is shown in Figure~\ref{figDAG}.
A good introduction to directed acyclic graphs is given by
\citet{Pea88}.

%s3.1 #&#
\subsection{Fibres}
Henceforth let $\mathbf{F} = \{F_1,\ldots,F_k\}$ denote $k$ random fibres.
As outlined earlier and illustrated in
Figure~\ref{figFibreParameters}, the fibre $F_j$ is determined by a
reference point $\omega_j$ and arc lengths $l_{j,1},l_{j,2}$.
It is also written $F_j = F_j(\omega_j,l_{j}, \nu_{\mathrm{FO}} )$
[where $l_j=(l_{j,1},l_{j,2})$] to indicate that it is
a deterministic function of $\omega_j$ and $l_{j}$ once $\nu_{\mathrm
{FO}}$ is given.
For the list of reference points we write $\bolds\omega= \{\omega
_1,\ldots,\omega_k\}$, and the arc length vectors are given by $\mathbf{l}
= \{l_1,\ldots,l_k\}$.
We use $l_{j,T} = l_{j,1} +l_{j,2}$ as a shorthand for the total arc
length of the $j$th fibre.
Note that in general the orientation field $\upsilon_{\mathrm{FO}}$ may
possess singularities, which would constrain the choice of the lengths
$l_j=(l_{j,1},l_{j,2})$;
however, this does not arise in our examples.

%s3.2 #&#
\subsection{Signal points}
Points from the observed pattern may be either signal or noise. Signal
points are typically clustered around fibres.
The model we use assigns an anchor point $p_i$ on some fibre to each
data point $y_i$.
The data point is then displaced from $p_i$ by an isotropic bivariate
normal distribution [i.e., $y_i \sim\operatorname{MVN}(p_i,\sigma
_{\mathrm
{disp}}^2\mathbf{I}_2)$, where $\mathbf{I}_2$ is the $2\times2$
identity matrix].

The fibre on which $p_i$ is located is determined by an auxiliary
variable $X_i$, so $X_i=j$ if and only if $p_i\in F_j$.
The $p_i$'s on the $j$th fibre are spaced such that the vector of
arc-length distances between adjacent points is proportional to a
Dirichlet distributed random variable.
Setting an appropriate parameter for the Dirichlet distribution will
encourage points to be either evenly spread, clustered along the fibre,
or placed independently at random along the fibre.

The probability that point $y_i$ is allocated to the $j$th fibre ($X_i
= j$) is proportional to the total length of fibre $F_j$.
This ensures that the mean points per unit streamline remains
approximately constant.

%f3 #&#
%

%s3.3 #&#
\subsection{Noise points}
Noise is then added as a homogeneous Poisson process.
This is included in the model by first allocating each point $y_i$ to
noise or signal (stored in auxiliary variable $Z_i = 1$ or $0$ for
signal or noise, resp.).
Point $y_i$ is allocated to signal independently of the allocations of
all other points.
The prior probability that $y_i$ is allocated to signal is given by
$\varepsilon_i$.
If the point is signal, then its location is distributed as outlined in
the previous subsection.
Otherwise, if the point is noise, it is distributed uniformly across
the window $W$.

%s3.4 #&#
\subsection{Total number of points}

The total number of points $m$ is assumed to be Poisson distributed.
The mean total number of points $\mu_{\mathrm{total}}$ is defined to be
equal to some function of $\mu_{\mathrm{signal}}$, the mean number of
signal points, and $\rho$, a parameter governing the number of noise points.
For the sake of simplicity we set $\rho$ to be the prior assumption on
the proportion of the total points that are noise points and define $\mu
_{\mathrm{total}} = \mu_{\mathrm{signal}} / (1-\rho)$.
The mean number of signal points $\mu_{\mathrm{signal}}$ is assumed to
be proportional to the total sum of the fibre arc lengths.
Hence, $m$ is assumed to be Poisson distributed with mean
%
%e1 #&#
%
\begin{equation}
\mu_{\mathrm{total}} = \Biggl(\sum_{j=1}^k (l_{j,T})
\Biggr)\frac{\eta}{1-\rho},
\end{equation}
where $\rho= \beta_{\mathrm{signal}}/({\alpha_{\mathrm{signal}}+\beta
_{\mathrm{signal}}})$ is the prior
estimate of the proportion of points that are signal and $\eta$ is a
density parameter.

The assumption that the mean number of noise points is proportional to
the mean total number of points (and the
fibre length) is particularly well suited to the fingerprint example
[see Section~\ref{secfingerprint}], where
noise points arise as artifacts of the pore detection process along the
fingerprint ridges.
Implementation of
alternative relationships
between the mean number of signal and noise points would be a
straightforward matter.

%s3.5 #&#
\subsection{Priors}

In the examples given in the next Section~\ref{secexamples} we use the
following priors:
\begin{eqnarray}
P(\mathbf{l}|k, \lambda)
&=&
\prod_{j=1}^k P({l_{j,1}}| \lambda)P({l_{j,2}}| \lambda)
\qquad\mbox{where } {l_{j,\cdot}} \sim\operatorname{Exp}(1/\lambda), \nonumber\\[-2pt]
P(\bolds\omega|k)
&=&
\prod_{j=1}^k P(\omega_j) \qquad\mbox{where } \omega_j\sim
\operatorname{Uniform}(W), \nonumber\\[-2pt]
P(k|\kappa) &\sim& \operatorname{Poisson}(\kappa), \nonumber\\[-2pt]
%P(m|\mathbf{l},\eta) &\sim& \textrm{Poisson}((\sum_{j=1}^k
%(l_{j,T}))\eta\frac{\alpha_{\mathrm{signal}}+\beta_{
P(\bolds{\epsilon}|\alpha_{\mathrm{signal}}, \beta_{\mathrm{signal}})
&=& \prod_{i=1}^m P(\epsilon_i|\alpha_{\mathrm{signal}}, \beta_{\mathrm
{signal}}) \nonumber\\[-2pt]
&&\eqntext{\mbox{where } \epsilon_i \sim\operatorname{Beta}(\alpha_{\mathrm
{signal}}, \beta_{\mathrm{signal}}).}
\end{eqnarray}
Here $m$ is the total number of points in $\{y_1,\ldots,y_m\}$.

The above prior models are common, parsimonious choices that appear
flexible enough for a range of applications including the examples
considered in Section~\ref{secexamples}. However, if
application-specific prior information suggests alternative prior
models, then these can be accommodated in the presented framework.

%s3.6 #&#
\subsection{Posterior}
We are interested in the posterior distribution of fibres (and various
other parameters) given a particular instance of the point process.
This posterior is given by
%
%e2 #&#
%
\begin{eqnarray}
\label{eqBayesRule}
&&
\pi(\mathbf{F}, \mathbf{l}, \bolds\omega, k, \upsilon_{\mathrm
{FO}}, \bolds{\epsilon},\mathbf{Z}, \mathbf{X},\mathbf{p})
\nonumber\\
&&\qquad= P(\mathbf{F}, \mathbf{l}, \bolds\omega, k, \upsilon_{\mathrm
{FO}}, \bolds{\epsilon},\mathbf{Z}, \mathbf{X},\mathbf{p}
%, m
|\mathbf{y}) \nonumber\\
&&\qquad\propto P(\mathbf{F}, \mathbf{l}, \bolds\omega, k, \upsilon
_{\mathrm{FO}}, \bolds{\epsilon},\mathbf{Z}, \mathbf{X},\mathbf{p})
%,m
\nonumber\\[-8pt]\\[-8pt]
&&\qquad\quad{} \times
L(\mathbf{F}, \mathbf{l},\bolds\omega, k, \upsilon_{\mathrm{FO}},
\bolds{\epsilon},\mathbf{Z}, \mathbf{X},\mathbf{p}
%,m
|\mathbf{y}) \nonumber\\
&&\qquad=
P(\mathbf{l}|k)
P(\bolds\omega|k)
P(k)
P(\upsilon_{\mathrm{FO}})
%P(m|\mathbf{l})
P(\bolds{\epsilon}
%|m
)
P(\mathbf{Z}|\bolds{\epsilon})
\nonumber\\
&&\qquad\quad{} \times
P(\mathbf{X}|\mathbf{Z}, \mathbf{l})
P(\mathbf{p}|\mathbf{F},\mathbf{X})
L(\mathbf{F}, \mathbf{l}, \bolds\omega, k, \upsilon_{\mathrm{FO}},
\bolds{\epsilon},\mathbf{Z}, \mathbf{X},\mathbf{p}
%,m
|\mathbf{y}).\nonumber
\end{eqnarray}
Here $P(\cdot)$ indicates a prior distribution.
We omit $P(\mathbf{F}|\mathbf{l_1}, \mathbf{l_2}, \bolds\omega,
\upsilon_{\mathrm{FO}})$, as it is deterministically calculated.

Section~\ref{secBDMCMC} describes how to sample from this posterior
distribution using Markov chain Monte Carlo techniques.

%s3.7 #&#
\subsection{Computational simplifications}
Computer implementation makes it necessary to
% discretize the infinite dimensional
represent the
field of orientations
by a discrete structure.
We adopt the simple approach of estimating the field of orientations at
a dense regular grid of points over $W$.
Integral curves are calculated stepwise by estimating the orientation
at a point by its value at the nearest evaluated grid point and extending
the curve
a small distance in that direction.
Note that the choice of direction (from the two available for each
orientation) is made so that the angle between adjacent linear segments
is greater than~$\pi/2$.

Consequently, fibres are stored as piecewise-linear curves and further
calculations are performed on these approximations.
Of course,
% However we can take
this discretization
% to be arbitrarily small
can be arbitrarily reduced
(at a
%correspondingly large
computational cost) to improve the accuracy of the approximation.

%s4 #&#
\section{Construction of field of orientations}
\label{secgradfield}
We must of course identify a~meth\-od for calculating the field of orientations.
% We will derive great
It is
computationally advantageous
% benefit from producing
to generate
a field of orientations which is likely to contain (be integrated by)
fibres that fit the data well (produce a high likelihood).
The most natural way to do this is to base the calculation of the field
of orientations on the data,
using
% in
an empirical Bayes technique.
The use of empirical Bayes to find the prior for the field of
orientations distribution means that aspects of the prior, or
parameters of the prior, are estimated from the data.

An alternative approach would be to use a fully Bayesian model, where
we would treat the field of orientations as an independent random
variable~$\Upsilon_{\mathrm{FO}}$.
We would then need to identify its state space and a corresponding
$\sigma$-algebra, transition kernel and prior on this state space.
These could be derived from random field theory [see, e.g., \citet
{Adl07}], using an appropriate covariance function to maintain
smoothness in the field of orientations, however, there are a number of
issues with this approach.
In practice,
one may expect the task of
sampling a random field of orientations to be computationally
expensive, particularly if the covariance function does not have a
simple form (as is likely in this model).
Calculations relating to the conditional distribution\vadjust{\goodbreak} of the field of
orientations given the fibres are likely to lead to unfeasible
computational complexity.
A~further issue is that this approach leads to a huge space of possible fibres,
resulting in corresponding difficulties in ensuring this space is
properly explored. Use of the information given in the data will help
to limit this space to a more easily explorable restricted class of
suitable fields of orientations.

Here we use the data to make local orientation estimates and smooth
these to produce a field of orientations estimator.
As we are specifically interested in orientation estimates arising from
the signal data, we can choose to weight the contribution of each point
to the field of orientations estimator by how likely it is to be noise
or signal.

Estimation of a field of orientations given that $y_1,\ldots,y_m$ are all
signal points is outlined in the following section. Section \ref
{secgradfieldeps} shows how to extend
this to take account of the information given in the vector of
probabilities that points are signal $(\varepsilon_1,\varepsilon
_2,\ldots,\varepsilon_m)$.\vspace*{-3pt}

%s4.1 #&#
\subsection{Estimation for all signal points}
\label{secgradfieldall}

The mapping and tensor method described in \citet{Tho08} [and further
discussed in \citet{Su09}] is applied to the point pattern to construct
a tensor at each point.
To this we apply a~Gaussian kernel smoothing in the log-Euclidean
metric to construct a~tensor field.
The tensor field is represented by an assignation to each point of
a~$2\times2$ nonnegative definite matrix whose principal eigenvector
indicates the dominant orientation at that point; the relative
magnitude of the eigenvalues indicates the strength of the dominant orientation.
The field of orientations
% is constructed as
assigns the orientation of this principal eigenvector to each
respective point.
If the principal eigenvector is not unique at a certain point (which is
to say that the eigenvalues are equal there),
then that indicates a singularity in the field of orientations.

Three-dimensional tensor fields of this kind are commonly used in
diffusion tensor imaging (DTI) to understand brain pathologies such as
multiple sclerosis, schizophrenia and strokes. DTI is used to analyze
images of the brain collected from magnetic resonance imaging (MRI)
machines. The MRI scan detects diffusion of water molecules in the
brain and uses the data to infer the tissue structure that limits water
flow. The three-dimensional diffusion tensor describes the orientation
dependence of the diffusion. Roughly speaking, the eigenvalues indicate
a measure of the proportion of water molecules flowing in the
associated eigenvector direction. For more information on DTI see, for
example, \citet{Bih01} and \citet{Chu07}.

Let $y_1,\ldots,y_m$ denote the spatial data points.
A tensor is constructed at a point $y_j$ using a nonlinear
transformation applied to the vectors $v^i = (v_1^i,v_2^i) =
\overrightarrow{y_jy_i}$ for $i\neq j$ [\citet{Tho08},
\citet{Su09}].
Specifically,
%
%e3 #&#
%
\begin{equation}
\tilde{v}^i = (\tilde{v}^i_1,\tilde{v}^i_2) = \exp
\biggl(-\frac{((v_1^i)^2+(v_2^i)^2)}{2\sigma_{\mathrm
{FO}}^2}\biggr)
\frac{(v_1^i,v_2^i)}{\sqrt{(v_1^i)^2+(v_2^i)^2}},
\end{equation}
where $\sigma_{\mathrm{FO}}$ is a scaling parameter.\vadjust{\goodbreak}

The tensor at $y_j$ is then represented by
%
%e4 #&#
%
\begin{equation}\label{eqtensor}
T_0(y_j) = \sum_{i\neq j} (\tilde{v}_1^i, \tilde
{v}_2^i)^{\mathrm{T}}(\tilde{v}_1^i, \tilde{v}_2^i).
\end{equation}

The result of the above method is to produce a set of $2\times2$
matrices located over a sparse set of locations.
In order to create a field of orientations, we must then interpolate to
get a tensor field.
Thus, we use the orientation of the principal eigenvector, where
defined, to construct a field of orientations.

Interpolation of tensors inevitably requires a notion of tensor metric.
We elect to work in the log-Euclidean metric [see \citet{Ars06}].
For an extended account of tensor metrics see \citet{Dry08}.
Log-Euclidean calculations are simply Euclidean calculations on the
tensor logarithms which are transformed back to tensor space by taking
the exponential.
The tensors arising in this study can all be represented by positive
definite matrices.
Tensor logarithms are therefore well defined as logarithms of these matrices.
However, the matrix calculated in (\ref{eqtensor}) will have a
zero-eigenvalue if the points are collinear, and therefore not be
positive definite.
If all points are truly collinear, then our approach breaks down---and
indeed the method is not intended for such noise-free data sets.
The more common situation is that one vector $\tilde{v}^i$ dominates
the tensor representation as calculated in (\ref{eqtensor})
due to the relative distances between points.
Typically this occurs if two points are close while other points are
far from the pair.
Due to rounding errors, the contribution of other points to the matrix
becomes zero, and the two remaining points are collinear by definition.
In order to avoid an error in the log-Euclidean calculation, if a
tensor has at least one zero-eigenvalue, then it is replaced by the
``uninformative'' identity matrix, suggesting a lack of directional
information. Thus, we take a conservative approach that
excludes any potentially misleading directional
information.\looseness=1

We calculate the interpolated tensor field $T_{h_{\mathrm{FO}}}(x)$ for
$(x\in W)$ as a kernel smoothing procedure, using a Gaussian kernel
\(f\) with variance parameter~$h_{\mathrm{FO}}^2$ in the log-Euclidean metric.
Hence, when the smoothing parameter~\(h_{\mathrm{FO}}\) is
positive,~$h_{\mathrm{FO}}>0$,
% \begin{eqnarray} \label{LEMeanf}
%
%e5 #&#
%
\begin{equation} \label{LEMeanf}
T_{h_{\mathrm{FO}}}(x) =
% &=&
% \exp(\frac{\int_W f(\textrm{dist}(x,z)) \log(T_0(z))\d z}{
% \nonumber\\
% &=&
\exp\biggl(\frac{\sum_{y_i\in\{y_1,\ldots,y_m\}} f(\operatorname{dist}(x,y_i))
\log(T_0(y_i))}{\sum_{y_i\in\{y_1,\ldots,y_m\}}f(\operatorname{dist}(x,y_i))}\biggr).
\end{equation}
%
% \end{eqnarray}
% as we set $\log(T_0(z)) = \underline{0}_2$ for $z\notin\{y_1,

The field of orientations $\upsilon_{FO}(y_1,\ldots,y_m;h_{\mathrm
{FO}})(x)$ for $x\in W$ is then defined to be equal to $\tan
^{-1}(v_1(x)/v_2(x))$, where $(v_1(x),v_2(x))$ is the principal
eigenvector of the matrix representation of $T_{h_{\mathrm{FO}}}(x)$.

In most instances this procedure will give a good estimation of a
suitable field of orientations for modeling the point process with
integral fibres.
The smoothing method has the drawback that it can create a bias around
areas of high curvature (rapidly varying orientation) in the field of
orientations.
Potential solutions have been analyzed and found to perform
well.\vadjust{\goodbreak}
The magnitude of the bias was found to be proportional to the smoothing
parame\-ter~$h_{\mathrm{FO}}$, hence, these solutions typically involve
compensating for~$h_{\mathrm{FO}}$: we give two examples:
\begin{longlist}[(B)]
\item[(A)] We allow the smoothing parameter to vary over the window,
such that lower values are used in areas of high point intensity.
This ensures that less information is extrapolated to regions where
there is already sufficient data, and therefore less bias occurs in
those regions.
\item[(B)] We consider two instances of the field of orientations with
different values of smoothing parameter $h_{\mathrm{FO}}'$ and
$h_{\mathrm{FO}}''$; the unbiased estimate can be found by
extrapolating back to estimate the field of orientations with
$h_{\mathrm{FO}}=0$.
\end{longlist}
We do not go into further detail here because doing so would distract
from the main ideas in this paper, but we do notice a minor effect of
this bias in the examples in Section~\ref{secexamples}.
Details of these bias corrections can be found in \citet{Hill-2011}.

%s4.2 #&#
\subsection{Estimation using signal probabilities}
\label{secgradfieldeps}

We extend this field of orientations estimation to take account of the
vector of probabilities that points are signal $(\varepsilon
_1,\varepsilon_2,\ldots,\varepsilon_m)$ by weighting the construction of
the initial tensor and also weighting the contribution of each initial
tensor to the kernel smoothing.

Specifically, the initial tensors are represented by
%
%e6 #&#
%
\begin{equation}
T_0(y_j) = \sum_{i\neq j} (\tilde{v}_1^i, \tilde
{v}_2^i)^{\mathrm{T}}(\tilde{v}_1^i, \tilde{v}_2^i)\epsilon_i
\end{equation}
for each point $y_j$, and the tensor field becomes
%
%e7 #&#
%
\begin{equation}\label{eqtensorfield}
T_{h_{\mathrm{FO}}}(x) =
\exp\biggl(\frac{\sum_{y_i\in\{y_1,\ldots,y_m\}} \epsilon_i f
(\operatorname{dist}(x,y_i)) \log(T_0(y_i))}{\sum_{y_i\in\{y_1,\ldots
,y_m\}} \epsilon_i
f(\operatorname{dist}(x,y_i))}\biggr).
\end{equation}

This weighting allows points that are more likely to be signal points
to have a greater effect on the field of orientations estimation.
As $\varepsilon_i\rightarrow0$ the effect of the point \(y_i\) on the
field of orientations tends to zero, whereas if \mbox{$\varepsilon_i=1$} for
all $i$ we would be performing the calculation described in Section~\ref
{secgradfieldall}.

%s5 #&#
\section{Sampling from the posterior distribution}
\label{secBDMCMC}

We seek to infer some characteristics of the fibre process when only
the point pattern is known.
Typical attributes of interest include the number of fibres, where they
are located/orientated, which points arose from which fibre and which
points arose from background noise.

Direct inference from the model is hindered by the complexity of its
hierarchical structure.
Hence, we choose to draw samples from the posterior distribution of the
fibres and other variables using Markov chain Monte Carlo methods.
Characteristics of interest can be estimated from these samples.

%s5.1 #&#
\subsection{Hyperparameters}
\label{sechyperparameters}

As a rough guideline, hyperparameters can be chosen as follows.

The prior mean number of fibres $\kappa$
%,
and the prior mean length of fibres $\lambda$ can be estimated from
any prior knowledge or expectations of the fibres.
The deviation of points from fibres $\sigma_{\mathrm{disp}}^2$ can be
estimated using prior knowledge of fibre widths and the approximation
that 95\% of points should lie within $2.45\sigma_{\mathrm{disp}}$ of
the center of a fibre.
The density of points per unit length of fibre $\eta$ can be similarly
estimated.

Orientation field parameters $h_{\mathrm{FO}}$ and $\sigma_{\mathrm
{FO}}$ should be chosen to ensure the orientation field is smooth.
These can be estimated by evaluating the orientation fields for
different selections of $h_{\mathrm{FO}}$, $\sigma_{\mathrm{FO}}$ and
choosing from this set.
If the proportion of noise points is approximately known, then the
hyperparameters $\alpha_{\mathrm{Signal}}$ and $\beta_{\mathrm
{Signal}}$ can be suitably estimated, however, we suggest choosing the
parameters such that $\alpha_{\mathrm{Signal}},\beta_{\mathrm
{Signal}}>1$ to ensure good mixing properties of the Markov chain Monte
Carlo sampling algorithm.
Otherwise the noise hyperparameters can be set equal to $1$, indicating
no prior knowledge.

Alternatively, if little prior information is known about the nature of
the latent curvilinear structure, then it would be feasible to extend
the empirical Bayes step to include the estimation of further prior parameters.

%s5.2 #&#
\subsection{Birth--death Monte Carlo}

The starting point for our algorithm is a continuous time birth--death
Markov chain Monte Carlo (BDMCMC) in which fibres are created and die
at random times controlled by predetermined or calculated rates. This
enables exploration of a wide range of models with different numbers of
fibres, and is suited to this type of clustered data. See
\citet{Mol04} for an introduction to spatial birth--death
processes. For the sake of brevity we present in the following the key
points of the algorithm. Further details can be found in
\citet{Hill-2011}.

Here we choose to fix the birth rate and calculate an appropriate death
rate at each step to maintain detailed balance.

Following a birth or death we update the following auxiliary variables:
$\mathbf{Z}$ to $\mathbf{Z}'$, the indicators of the components
(signal/noise) to which the points are associated;
$\mathbf{X}$ to $\mathbf{X}'$, the indicators of the fibres to which
the signal points are associated;
$\mathbf{p}$ to $\mathbf{p}'$, the vector $(p_1,\ldots,p_m)$ where $p_i$
is the point on the fibre to which the
%each
data point $y_i$ is associated.

%s5.3 #&#
\subsection{Births of fibres}
%The
Recall that the parameterization of fibres is described in Section \ref
{secmodelfibres}.
Birth events occur randomly at rate $\beta$.
Upon the occurrence of a birth, the number of fibres is updated from
$k$ to $k+1$, and a new fibre is introduced by sampling a reference
point $\omega_{k+1}$ and lengths $l_{k+1,1},l_{k+1,2}$ from the prior
distributions $P(\omega), P(l)$, respectively.
The new fibre $F_{k+1}$ is then calculated by integrating the field of
orientations according to these parameters, and the set of fibres
$\mathbf{F} = F_1,\ldots,F_k$ is updated to $\mathbf{F}' = F_1,\ldots
,F_k,F_{k+1}$.
In order to ensure that the distribution of the lengths
$l_{k+1,1},l_{k+1,2}$ is independent of the respective directions in
which the field of orientations is integrated, we choose them to be
independently and identically distributed.
%We choose the lengths $l_{k+1,1},l_{k+1,2}$ to be independently and
%identically distributed so that the choice of which length relates to
%which direction in the field of orientations integration is irrelevant.

Data points which are currently assigned to the noise component are
reassigned to noise or signal with proposal probability dependent on
the new fibre $Q_{\mathrm{birth}}(\mathbf{Z}\mapsto\mathbf{Z}'|
\mathbf{Z},\bolds\varepsilon, F_{k+1})$.

Finally, new values are proposed for all of $\mathbf{X}, \mathbf{p} $
according to a proposal density $Q(\mathbf{X}, \mathbf{p}\mapsto\mathbf
{X}', \mathbf{p}')$.
For simplicity, we choose not to sample from the full conditional
distribution of $\mathbf{p}$ and $\mathbf{X}$,
%in that we do not require the proposal density to account for the
%Dirichlet requirement on $\mathbf{p}$ or the distribution of
%allocations of points to fibres.
%Rather
but rather
%the auxiliary variables $\mathbf{p}$ and $\mathbf{X}$ are proposed
from a density proportional to the likelihood $ L(\mathbf{p}, \mathbf
{X}|\mathbf{y})$.

In full, the birth density of fibre $F_{k+1}$ including updates of
auxiliary variables to $\mathbf{Z}', \mathbf{X}', \mathbf{p}'$ is given by
%
%e8 #&#
%
\begin{eqnarray}
&&b(F_{k+1}, \omega_{k+1}, l_{1,{k+1}}, l_{2,{k+1}},\mathbf{Z}',
\mathbf{X}', \mathbf{p}') \nonumber\\
&&\qquad=\beta P(\omega_{k+1}) P(l_{1,{k+1}})P(l_{2,{k+1}})
Q_{\mathrm{birth}}(\mathbf{Z}\mapsto\mathbf{Z}'|\mathbf{Z},\bolds\epsilon,
F_{k+1})\\
&&\qquad\quad{}\times Q(\mathbf{X}, \mathbf{p}\mapsto\mathbf{X}', \mathbf{p}').\nonumber
\end{eqnarray}

%s5.4 #&#
\subsection{Deaths of fibres}

We must calculate the death rate $\delta_j$ for each fibre to ensure
detailed balance holds.
Following the death of fibre $F_j$, the variables~$\mathbf{F},
\bolds\omega$ and $\mathbf{l}$ are updated by omitting the $j$th term.
Further, auxiliary variables~$\mathbf{Z}, \mathbf{X}$ and $\mathbf{p}$
are all updated.
All points allocated to fibre $F_j$ are now allocated to noise.
We call this trivial proposal density $Q_{\mathrm{death}}(\mathbf{Z}'|
\mathbf{Z},\mathbf{X},j)$.
Again the final step is to propose new values for all of $\mathbf{X}$
and $\mathbf{p}$ according to proposal density $Q(\mathbf{X}, \mathbf
{p}\mapsto\mathbf{X}', \mathbf{p}')$.

Hence, the death rate that satisfies detailed balance for the $j$th
fibre is given by
\begin{eqnarray}
\delta_j & = &
\frac{P(\mathbf{l_1}\setminus l_{1,j}, \mathbf{l_2}\setminus l_{2,j}|k-1)}
{P(\mathbf{L_1}, \mathbf{L_2}|k)}
\frac{P(\bolds\omega\setminus\omega_j|k-1)}
{P(\bolds\omega|k)}
\frac{P(k-1)}
{P(k)}
\nonumber\\
& &{} \times
\frac{
b(F_j, \omega_j, l_{1,j}, l_{2,j},\mathbf{Z}', \mathbf{X}', \mathbf{p}')}
{Q_{\mathrm{death}}(\mathbf{Z}'|\mathbf{Z},\mathbf{X},j) Q(\mathbf{X}, \mathbf
{p}\mapsto\mathbf{X}', \mathbf{p}')}
\\
& = &
\frac{P(k-1)}
{P(k)}
\frac{
\beta Q_{\mathrm{birth}}(\mathbf{Z}'\mapsto\mathbf{Z}|\bolds\epsilon, F_{j})
Q(\mathbf{X}', \mathbf{p}'\mapsto\mathbf{X}, \mathbf{p})
}{
Q_{\mathrm{death}}(\mathbf{Z}'|\mathbf{Z},\mathbf{X},j) Q(\mathbf{X}, \mathbf
{p}\mapsto\mathbf{X}', \mathbf{p}')
}.\nonumber
\end{eqnarray}

%s5.5 #&#
\subsection{Additional moves}
\label{secaddMoves}

It can be desirable to add extra moves to the BDMCMC process to improve mixing.
Some possible moves which were all utilized in the examples in Section
\ref{secexamples} include the following:
\begin{itemize}
\item moving a fibre by a small amount (by perturbing the reference point),
\item resampling the lengths of a fibre (while keeping the reference
point fixed).\vadjust{\goodbreak}
%two.
\end{itemize}
Each of these events occur at some predefined rate, whence they are
proposed and either accepted or rejected according to the Metropolis
Hastings probability.

We may also wish to update other model variables, giving more flexibility
and improving the algorithm's exploration of the sample space.
The additional variable updates used in the examples in Section
\ref{secexamples}
include the following:
\begin{itemize}
\item proposing new signal-noise allocations of the data ($\mathbf{Z}$),
\item proposing new signal probabilities ($\bolds\varepsilon$)
according to nondegenerate
Beta distributions whose parameters depend on the current signal- noise
allocations $\mathbf{Z}$.
This move leads to an update in the prior for the field of orientations
due to the empirical Bayes step, hence, all fibres are resampled.
\end{itemize}
%
%The only hyperprior parameter which was updated in the examples was the
%constant of proportionality $\eta$ in the prior for the
%Poisson-distributed number of points.
%We chose a Gamma-distributed prior for $\eta$.
Details of all moves can be found in \citet{Hill-2011}.

Hyperprior parameters, such as the constant of proportionality $\eta$ in
the prior for the Poisson-distributed number of points or
$\sigma_{\mathrm{disp}}$ governing the deviation of points from fibres,
may also be updated.
We have chosen not to update any hyperprior parameters to reduce
complexity of the model.
%We may also wish to update other model variables, giving more
%flexibility and improving the algorithm's exploration of the %sample
%space.
%The additional variable updates used in the examples in Section
%this move leads to an update in the prior for the field %of
%orientations due to the empirical Bayes step, hence all fibres must be
%resampled;

%The only hyperprior parameter which was updated in the examples was
%the constant of proportionality $\eta$ in the prior for the
%%Poisson-distributed number of points.
%We chose a Gamma
%-distributed
% prior for $\eta$.

%s5.6 #&#
\subsection{Convergence and output analysis}

First recall that the signal probabilities $\bolds\varepsilon$ are
updated according
to nonsingular Beta distributions. Hence, the underlying tensor field
as defined in
(\ref{eqtensorfield}) will not become degenerate even when $\mathbf{Z}$
allocates all points to
noise.

Consider the set $A$ of states in which the fibre configuration is
empty and all points
are allocated to noise. In the following discussion we exclude any
degenerate states of equilibrium probability
zero. Inspection of the algorithm shows that the set $A$ can be reached
from any nondegenerate state
in finite time and so the birth--death process is $\phi$-irreducible.
Recurrence can be deduced
by noting that the set $A$ is visited infinitely often; see
\citet{KaspiMandelbaum-1994}.

We motivate a heuristic lower bound on a suitable burn-in time
by considering aspects of the prior derived after inspection of the
data (e.g., $\sigma_{\mathrm{disp}}, \lambda, \kappa$---see Section \ref
{sechyperparameters}), and estimating the number of fibre births
%of fibres which
that must occur before a fibre has been created around each potential
fibre cluster.
We approximate the lower bound by considering
%only
the number of fibre births required for
this to happen
%a fibre to be born
around the \textit{smallest}
suitable cluster of points.
%fibre cluster.

A lower bound on half the length of the shortest
suitable
%fibre
cluster is derived from the $10\%$ quantile of an exponentially
distributed random variable of rate $\kappa/\lambda$.
Then the probability that a point chosen at random from $W$ lies in a
region corresponding to an actual fibre of this length (up to $2\sigma
_{\mathrm{disp}}$ from the fibre) is approximated by
%
%e9 #&#
%
\begin{equation}
\frac{8\lambda\log(10/9)\sigma_{\mathrm{disp}}}{\kappa|W|}.
\end{equation}
%
%We estimate properties of the fibre set (e.g. number of fibres, spread
%of points) from the priors.
%In particular a suitable burn-in time will usually increase
%proportionally to the area of the window $|W|$.
%In particular we choose a burn-in time
%T_{\mathrm{burn}} = \min\{1500, \frac{|W|}{4 \sigma_{
%where $l_{\mathrm{short}}$ is an estimator of the length of the
%shortest fibre.
It follows that, with probability $0.99$, a fibre will be proposed in
the region corresponding to the shortest fibre within the first
%
%e10 #&#
%
\begin{equation}
\frac{\log(0.01)}
{\log(
1-{8\lambda\log(10/9)\sigma_{\mathrm{disp}}}/({\kappa|W|})
)}
\end{equation}
births.
Hence, we choose a burn-in time of
%
%e11 #&#
%
\begin{equation}
T_{\mathrm{burn}} =
\max\biggl\{1500,
\frac{\log(0.01)}
{\beta\log(
1-{8\lambda\log(10/9)\sigma_{\mathrm{disp}}}/({\kappa|W|})
)}
\biggr\},
\end{equation}
taking 1500 as a lower bound to ensure the burn-in time remains substantial.

Convergence was assessed by considering variables such as the number of
fibres $k$ or the number of noise points and using Geweke's spectral
density diagnostic; see \citet{Bro98}.
Convergence of a sequence of $n$ samples is rejected if the mean value
of the variable in the first $n/10$ samples is not sufficiently similar
to the mean value over the last $n/2$ samples.

We also tested convergence by assessing whether the mean sum of the
death rates is approximately equal to the birth rate $\beta$.
Consider $\delta_{\mathrm{total}}^kt^k$ where $\delta_{\mathrm
{total}}^k$ is the sum of the death rates of fibres after the $k$th
event (e.g., birth, death, etc.) and $t^k$ is the length of algorithmic
time before the next event.
If the MCMC has reached stationarity, then
%
%e12 #&#
%
\begin{equation}
Z_m =\frac{\sum_{k=1}^m\delta_{\mathrm{total}}^kt^k -m\beta/(2\beta
+r_{\mathrm{add}})}{\sigma_{\delta_{\mathrm{total}}t}\sqrt{m}}
\xrightarrow{D} N(0,1),
\end{equation}
where $\sigma_{\delta_{\mathrm{total}} t}$ is an estimate of the
standard deviation of $\delta_{\mathrm{total}}^kt^k$, $\beta$ is the
birth rate of fibres, and $r_{\mathrm{add}}$ is the sum of the rates of
any additional moves implemented (as\vspace*{1pt} suggested in Section
\ref{secaddMoves}). We used this result to test the convergence of
$1/m\sum_{k=1}^m\delta _{\mathrm{total}}^kt^k$ to
$\beta/(2\beta+r_{\mathrm{add}})$.

Bearing in mind the complexities of the underlying model, output
analysis showed no evidence for a lack of convergence.

%Convergence of the MCMC can be assessed by monitoring properties of
%the output samples, such as the number of fibres and the number of
%noise points.
%We can also compare the number of deaths and the number of births,
%confirming that they occur at the same rate.

Outputs of various variables are recorded at random times at some
constant rate.
The rate of this sampling (effectively the reciprocal of the thinning
of the Monte Carlo process) is chosen such that there is a low
probability that any of the fibres remain unchanged between samples.
The inclusion of the extra moves designed to improve mixing also helps
to decrease the thinning required.
The thinning is chosen approximately proportional to the number of
fibres (estimated based on aspects of priors derived
%following
from
inspection of the data).

%s6 #&#
\section{Simulation studies and applications}
\label{secexamples}

The implemented algorithm runs on a continuous time scale. Events occur
at a determined rate, either fixed or calculated to ensure that
detailed balance holds.
The units for the rate of an event are ``per unit of algorithm time.''
The BDMCMC is then allowed to run for\vadjust{\goodbreak} a large number of time units and
samples are taken at random times (at some fixed rate).
%Note that these rates (and notion of units of
Of course, the relationship of algorithm time
%) lie on a relative scale to the computer/processing time.
to actual processing time depends on hardware and implementation details.
Hardware details are described below.

In each of the following examples, the birth rate and the rate of other
moves (moving a fibre, adjusting lengths of a fibre, proposing a split
or a join, variable updates) were all unit rate.
The only exceptions were the signal probability ($\bolds\varepsilon
$) updates which were proposed at a rate of $0.1$ per unit of time, and
the recording of output variables at random times whose rate varied for
different data sets.

%We use the grid of pixels as the dense regular grid of points at which
We evaluate the field of orientations over a square grid of points,
each one unit length from its four nearest neighbors.
The total size of this grid is given by the dimensions of the window $W$.

All three examples were run on the cluster owned by the Statistics
Department in the University of Warwick using a Dell PowerEdge 1950
server with a 3.16~GHz Intel Xeon Harpertown (X5460) processor and 16
GB fully-buffered RAM.
The algorithm was implemented in Octave version 3.2.4.\footnote{The
Octave code for this algorithm is available at URL
\href{http://www2.warwick.ac.uk/go/ethonnes/fibres}{http://www2.warwick.ac.uk/}
\href{http://www2.warwick.ac.uk/go/ethonnes/fibres}{go/ethonnes/fibres}.}
The total run-times on the cluster ranged from 34.7 hours for the
fingerprint pore data (32,300 units of algorithm time) to 61.3 hours for
the earthquake data set (30,000 units of algorithm time).
Due to the limitations of the current version of Octave, the benefits
of a parallel implementation have not yet been explored.

Analysis has been performed on all four of the data sets shown in
Figure~\ref{figExample}.
However, for the sake of brevity, we omit discussion of results for the
first simulated point pattern [Figure~\ref{figExample}(a)] from this paper.

%s6.1 #&#
\subsection{Simulated example}

\begin{figure}
\begin{tabular}{@{}c@{\quad}c@{}}

\includegraphics{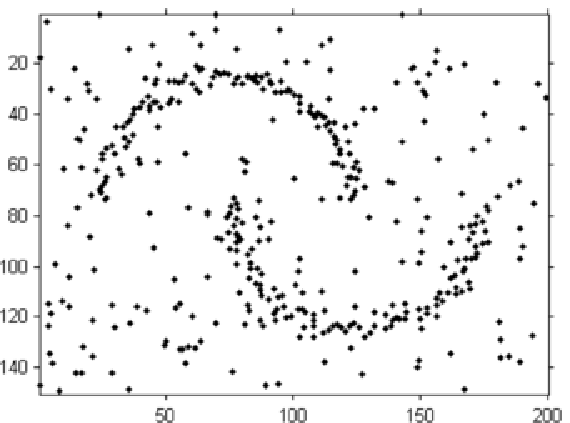}
 & \includegraphics{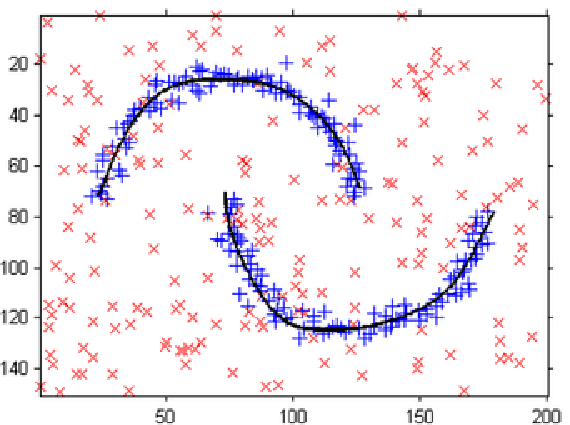}\\
(a) & (b)\\[4pt]

\includegraphics{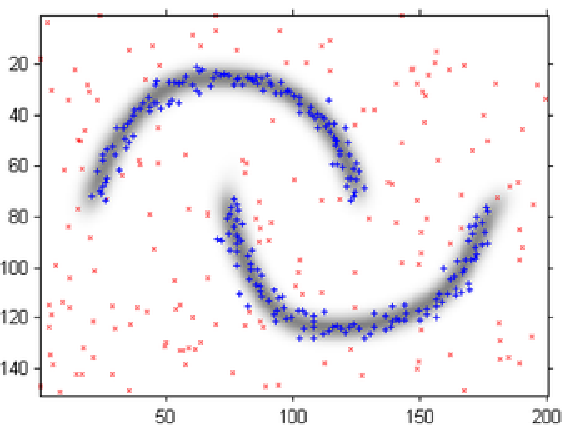}
 & \includegraphics{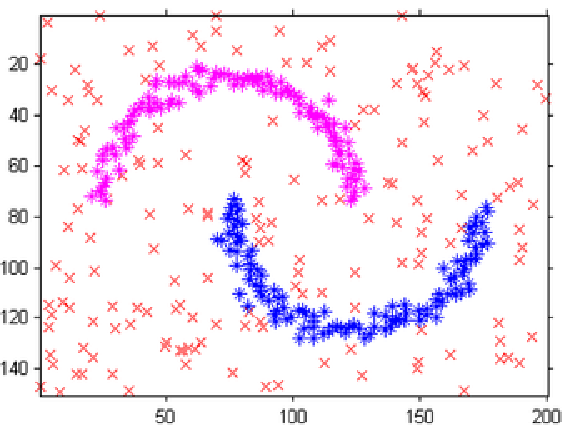}\\
(c) & (d)
\end{tabular}
\caption{Simulated example from Stanford and Raftery (\protect\citeyear{Sta00}).
\textup{(a)} Simulated data. \textup{(b)}~A~random sample from the
BDMCMC output. Fibres are represented by curves, pluses indicate points
allocated to signal and crosses indicate points allocated to noise in
this sample. \textup{(c)} Estimate of the density of signal points
found by smoothing a series of samples of fibres (darker areas indicate
higher densities). Pluses indicate points allocated to signal and
crosses indicate points allocated to noise in at least $50\%$ of
samples. The size of points representing the data has been reduced to
enhance the clarity of the density estimate. \textup{(d)}~Estimate of
the clustering of the signal points---different symbols indicate
different clusters, crosses indicate noise. Estimated by considering
how often pairs of points are associated with the same fibre across a
number of samples.}
\label{figeg2}
\end{figure}

Figure~\ref{figeg2}(a) shows the simulated data set used in
\citet{Sta00}. We include it here to facilitate comparison with
the methods proposed by Stanford and Raftery.
% show that our method performs equally well on the data set.
The data consists of 200 signal points and 200 noise points over a
$200\times150$ window, and is based on a family of two fibres each of
length 157.

%f4 #&#
%

%t1 #&#
%
\begin{table}
\caption{Results for Stanford and Raftery's simulated example: first
sub-table gives posterior~probabilities on the number of fibres, while
the second gives posterior means~and~50\% and~95\% HPD (highest
posterior density) intervals for a selection of~properties of the
posterior distribution conditional on the number of fibres. The
simulated data consists of~200 signal points and 200 noise points over
a $200\times150$ window, and is based on~a~family of two fibres each of
length 157. The dispersion parameter $\sigma_{\mathrm{disp}}$ is set to
3 and the prior mean probability that a point
is noise~is~$0.5$.~Posterior~probabilities only~given if nonzero to rounding error}
\label{tabeg2}
\begin{tabular*}{\tablewidth}{@{\extracolsep{\fill}}
lcd{3.2}cc@{}}
% & posterior mean & 50\% HPD interval & 95\% HPD interval \\
\hline
& \multicolumn{4}{@{}c@{}}{\textbf{Posterior probabilities for number of fibres}}\\
\hline
\multicolumn{2}{@{}l}{Number of fibres} & 2 & 3\hphantom{.00} & 4\hphantom{.00} \\
\multicolumn{2}{@{}l}{Posterior probability} & 0.73 & 0.23 & 0.04 \\
\hline
& \multicolumn{4}{@{}c@{}}{\textbf{Other properties conditioned on the number
of fibres}}\\[-4pt]
& \multicolumn{4}{c@{}}{\hrulefill}\\
& \multicolumn{1}{c}{\textbf{Number of}}
& \multicolumn{1}{c}{\textbf{Posterior}}
& \multicolumn{1}{c}{\textbf{50\% HPD}}
& \multicolumn{1}{c@{}}{\textbf{95\% HPD}}
\\
& \multicolumn{1}{c}{\textbf{fibres}}
& \multicolumn{1}{c}{\textbf{mean}} & \multicolumn{1}{c}{\textbf{interval}}
& \multicolumn{1}{c@{}}{\textbf{interval}}\\
\hline
Number of noise points &
2 & 181.85 & $[176,193]$ & $[156,205]$ \\
& 3 & 180.81 & $[179,197]$ & $[149,201]$ \\
& 4 & 178.78 & $[167,184]$ & $[155,197]$ \\
[4pt]
95th percentile of the distances
& 2 & 8.16 & $[7.17,8.32]$ & $[6.64,9.66]$ \\
\quad from signal points to fibres
& 3 & 8.06 & $[7.28,8.27]$ & $[6.61,9.53]$ \\
& 4 & 7.95 & $[7.20,7.91]$ & $[6.75,9.50]$ \\
[4pt]
Total length of fibres
& 2 & 317.68 & $[319,325]$ & $[301,325]$ \\
& 3 & 319.60 & $[315,322]$ & $[300,342]$ \\
& 4 & 320.10 & $[325,303]$ & $[303,325]$ \\
\hline
\end{tabular*}
\end{table}

%}
%fibres and circles indicate points %allocated to noise in this sample]{
%}
%a series of samples of fibres %(darker areas indicate higher
%densities)]{
%}
%symbols indicate different clusters, %dots indicate noise. Estimated
%by considering how often pairs of points are associated with the same
%fibre %across a number of samples.]{
%}

The birth--death MCMC was run for 60,000 units of algorithm time, the
first 30,000 of which were discarded.
Samples were taken at a rate of $0.033$ per unit of time.
The initial state was a randomly sampled set of $\kappa=2$ fibres.
Other hyperparameters were chosen as follows:
dispersion parameter $\sigma_{\mathrm{disp}} = 3$;
signal probability hyperparameters $\alpha_{\mathrm{signal}} = 1$ and
$\beta_{\mathrm{signal}} = 1$;
density parameter $\eta= 0.64$;
mean half-fibre length $\lambda= 78.5$; and the
Dirichlet parameter $\alpha_{\mathrm{Dir}} = 1.5$.

Figure~\ref{figeg2}(b)--(d) shows
that our model fits the data very well, albeit with a slight
extrapolation of fibres
beyond the curves used to generate the data set.
%we have chosen a reasonable model for the data.
The two fibres in the sample in Figure~\ref{figeg2}(b) compare
favorably with the principal curves fitted in \citet{Sta00}.\vadjust{\goodbreak}

%As in the previous example, we estimate properties of the posterior
%distribution.
%Posterior means and highest posterior density intervals are given in
%Table~\ref{tabeg2}.
Table~\ref{tabeg2} gives the posterior probabilities of the number of
fibres and the means and highest posterior density intervals of a
variety of properties conditional on the number of fibres.
The number of fibres is simply a count of the fibres present in each
sample; in this example we expect it to be around~2.
The number of points assigned to the noise component will typically be
closely correlated with the number of fibres.
With more fibres comes a greater chance of there being a fibre close to
a given point and hence a~greater chance that it is a signal point.
We take the 95th percentile of the distances of signal points from
their associated points on fibres for each sample.
This summarizes the dispersion of points from the fibres.
It is comparable to $2.45\sigma_{\mathrm{disp}}$, where $\sigma
_{\mathrm{disp}}$ is the dispersion parameter (set to $3$ in this
example).
The constant $2.45$
%arises as square root of the 95th percentile of a chi-squared
%distribution, and hence
arises as the 95th percentile of the Euclidean distance from the
origin to a bivariate standard-normal distributed random variable. The
curvature bias in the field of orientations results in a mild bias on
the 95th percentile of distances from signal points to anchor points.

In this example, less points are associated to noise than were
simulated as noise points in the data
generation. This is partly due to the high intensity of noise points,
and also explained by a slight bias in the length of the fibres. The
posterior statistics on the lengths of the fibres
suggest that the extension of fibres beyond their known length (of
$157$) is supported by the high intensity of noise points.
This extrapolation is sometimes beneficial, particularly for fibre
reconstruction
in areas of missing data.
Here the extrapolation is less desirable, as it suggests there is
evidence for fibres in the background noise.

The extrapolation of fibres into less dense regions of points can be
reduced by choosing a higher Dirichlet parameter $\alpha_{\mathrm
{Dir}}$ for the distribution of anchor points along the fibres.
This decreases the posterior density of fibres lying through point
clusters of nonconstant
intensity.
%Unfortunately
%However, as no proposals account for the distribution of anchor points
%along fibres, increasing $\alpha_{\mathrm{Dir}}$ leads to an anchor
%point distribution with higher modal peaks, and there proposed moves
%are more frequently rejected.
The drawback of increasing~$\alpha_{\mathrm{Dir}}$ is that
%proposed moves are more frequently rejected.
%This is because (A) no proposals account for the distribution of
%anchor points along fibres, and (B)
a large value of $\alpha_{\mathrm{Dir}}$ leads to a multimodal anchor
point distribution with most of the probability mass concentrated at
the modes. As the proposal of a birth/change of a fibre does not take
account of the shape
of anchor point distribution, the
%weighted around the modes,
proposal of a state with low posterior density is more likely for
larger $\alpha_{\mathrm{Dir}}$.

\subsection{Application: Earthquakes on the new Madrid fault line}

The epicenters of earthquakes along seismic faults are a good example
of point data clustered around a system of fibres with additional
background noise.
Here the fibres are the unknown fault lines.
\citet{Sta00} consider the structure of the data set of earthquakes
around the New Madrid fault line in central USA.
We use data on earthquakes in the New Madrid region between 1st Jan
2006 and 3rd Aug 2008 (inclusive) taken from the CERI (Center for
Earthquake Research and Information) found at
\href{http://www.ceri.memphis.edu/seismic/catalogs/cat\_nm.html}{http://www.ceri.}
\href{http://www.ceri.memphis.edu/seismic/catalogs/cat\_nm.html}{memphis.edu/seismic/catalogs/cat\_nm.html}.

%}
%fibres and circles indicate points %allocated to noise in this sample]{
%}
%a series of samples of fibres %(darker areas indicate higher
%densities).
%The data is represented by small points
%The size of points representing the data has been reduced to enhance
%the clarity of the density estimate.]{
%}
%symbols indicate different clusters, %dots indicate noise. Estimated
%by considering how often pairs of points are associated with the same
%fibre %across a number of samples.]{
%}

The birth--death MCMC was run for 40,000 units of algorithm time, the
first 10,000 of which were discarded.
Samples were taken at a rate of $0.0167$ per unit of time.
The initial state was a randomly sampled set of $\kappa=4$ fibres.
Other hyperparameters were chosen as follows:
dispersion parameter $\sigma_{\mathrm{disp}} = 2$;
signal probability hyperparameters $\alpha_{\mathrm{signal}} = 4$ and
$\beta_{\mathrm{signal}} = 1$;
density parameter $\eta= 1.06$;
mean half-fibre length $\lambda= 30$; and the
Dirichlet parameter $\alpha_{\mathrm{Dir}} = 1.5$.

Table~\ref{tabeg3} gives some numerical properties of the posterior
distribution of fibres.

% data from load d:\Bryony\Buster_Output\ctOut_E_c40350.mat
%86.9 hours
%
%t2 #&#
%
\begin{table}
\caption{Results for earthquake data: first sub-table gives posterior
probabilities on the number of fibres, while the second gives posterior
means and 50\% and 95\% HPD (highest posterior density) intervals for a
selection of properties of the posterior distribution conditional on
the number of fibres. The data are all the recorded earthquakes in the
New Madrid region between 1st Jan 2006 and 3rd~Aug~2008; the data were
acquired from the CERI (Center for Earthquake Research and Information)
found at \protect\href{http://www.ceri.memphis.edu/seismic/catalogs/cat\_nm.html}{http://www.ceri.memphis.edu/}
\href{http://www.ceri.memphis.edu/seismic/catalogs/cat\_nm.html}{seismic/catalogs/cat\_nm.html}.
In total there are 317 points in a $300\times300$ window, the
dispersion parameter $\sigma_{\mathrm{disp}}$ is set to 2 and the prior
mean probability that a point is noise~is~$0.2$.
Posterior probabilities only given if nonzero to rounding error}
\label{tabeg3}
\begin{tabular*}{\tablewidth}{@{\extracolsep{\fill}}
lcd{3.2}cc@{}}
% & posterior mean & 50\% HPD interval & 95\% HPD interval \\
\hline
& \multicolumn{4}{@{}c@{}}{\textbf{Posterior probabilities for number of fibres}}\\
\hline
\multicolumn{2}{@{}l}{Number of fibres} & 6 & 7\hphantom{.00} & 8\hphantom{.00} \\
\multicolumn{2}{@{}l}{Posterior probability} & 0.56 & 0.36 & 0.07 \\
\hline
& \multicolumn{4}{@{}c@{}}{\textbf{Other properties conditioned on the number
of fibres}}\\[-4pt]
& \multicolumn{4}{c@{}}{\hrulefill}\\
& \multicolumn{1}{c}{\textbf{Number of}}
& \multicolumn{1}{c}{\textbf{Posterior}}
& \multicolumn{1}{c}{\textbf{50\% HPD}}
& \multicolumn{1}{c@{}}{\textbf{95\% HPD}}
\\
& \multicolumn{1}{c}{\textbf{fibres}}
& \multicolumn{1}{c}{\textbf{mean}} & \multicolumn{1}{c}{\textbf{interval}}
& \multicolumn{1}{c@{}}{\textbf{interval}}\\
\hline
Number of noise points &
6 & 42.99 & $[41,44]$ & $[38,48]$ \\
& 7 & 40.65 & $[38,41]$ & $[36,45]$ \\
& 8 & 42.06 & $[42,44]$ & $[35,45]$ \\
[4pt]
95th percentile of the distances
& 6 & 5.25 & $[4.94,5.19]$ & $[4.96,5.60]$ \\
\quad from signal points to fibres
& 7 & 5.29 & $[5.12,5.40]$ & $[4.85,5.89]$ \\
& 8 & 5.17 & $[4.94,5.22]$ & $[4.74,5.65]$ \\
[4pt]
Total length of fibres
& 6 & 257.24 & $[247,257]$ & $[246,269]$ \\
& 7 & 257.43 & $[257,262]$ & $[249,264]$ \\
& 8 & 252.89 & $[250,252]$ & $[248,265]$ \\
\hline
\end{tabular*}   
\end{table}

%f5 #&#
%
\begin{figure}
\begin{tabular}{@{}c@{\quad}c@{}}

\includegraphics{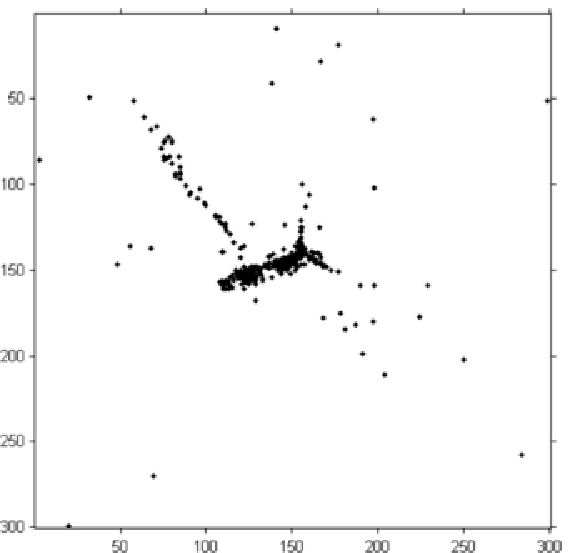}
 & \includegraphics{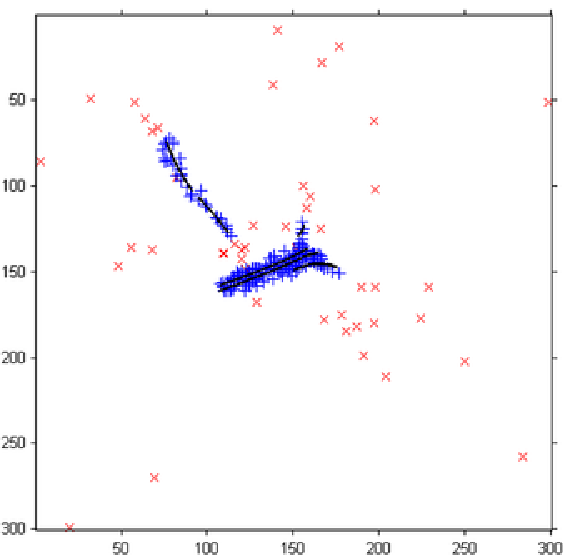}\\
(a) & (b)\\[4pt]

\includegraphics{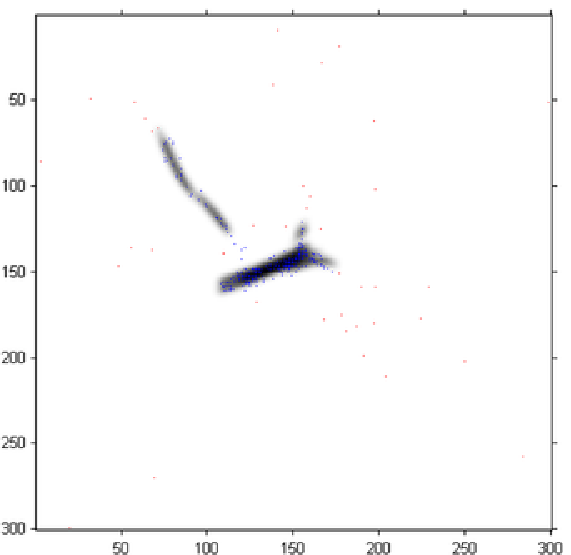}
 & \includegraphics{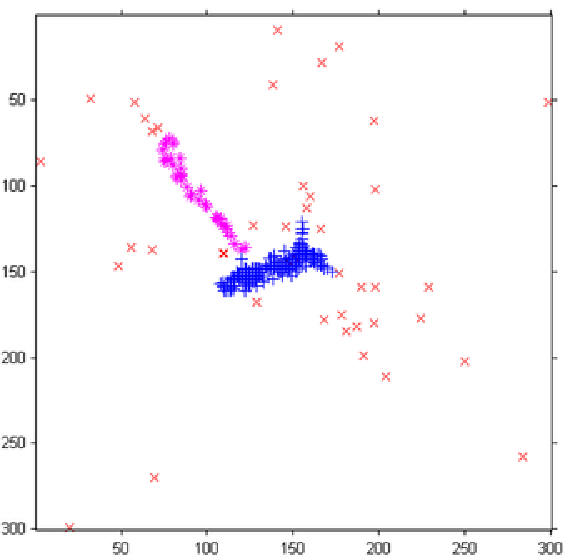}\\
(c) & (d)
\end{tabular}
\caption{New Madrid fault earthquake data. \textup{(a)} Earthquake
data. \textup{(b)} A random sample from the BDMCMC output. Fibres are
represented by curves, pluses indicate points allocated to signal and
crosses indicate points allocated to noise in this sample. \textup{(c)}
Estimate of the density of signal points found by smoothing a series of
samples of fibres (darker areas indicate higher densities). Pluses
indicate points allocated to signal in at least $50\%$ of samples. The
size of points representing the data has been reduced to enhance the
clarity of the density estimate. \textup{(d)} Estimate of the
clustering of the signal points---different symbols indicate different
clusters, crosses indicate noise. Estimated by considering how often
pairs of points are associated with the same fibre across a number of
samples.} \label{figeg3}
\end{figure}

Our method has the advantage over \citet{Sta00}, in that it does not try
to over-fit the fibres where there is less data.
Rather it uses information from surrounding data to extrapolate fibres
as required.
One limitation of our model is that every fibre is assumed to share a
number of properties.
In particular, the displacement of points from fibres (effectively the
width of influence of a fibre) and the intensity of signal points per
unit length of fibre are assumed to be constant, independent of the fibre.
These assumptions are not reasonable for this data as the ``thickness''
and density of points varies considerably.
This is apparent
%in Table~\ref{tabeg3} where the HPD intervals for the 95th percentile
%of the distance from points to fibres lie significantly above $2
in Figure~\ref{figeg3}(b) where
the central dense cluster
is described by
multiple parallel fibres.
% are required to
% explain.
The dispersion parameter $\sigma_{\mathrm{disp}}$ was chosen by
considering the apparent ``width'' of the longer thinner fibre, hence,
points around the shorter, wider fibre
effectively increase the 95th percentile of the point to fibre
distances, as given in Table~\ref{tabeg3}.
To overcome this issue, one could extend the model to
%In order to overcome this issue it would be necessary to
% extend our model to
allow different hyperparameters for each fibre.

%The curvature bias is evident in Figure~\ref{subeg3c} where the
%density estimate indicates that the long fibre has been extrapolated
%at an unexpected angle.
%This is because the high density of points around the shorter fibre
%dominate in the kernel smoothing, creating a bias in the field of
%orientations.

While multiple fibres in the central cluster is a common feature in
samples from this BDMCMC, Figure~\ref{figeg3}(d) indicates that the
agglomerative clustering algorithm identifies the points as arising
from the same cluster.

Interestingly, the total length of fibres does not appear to be
positively correlated to the number\vadjust{\goodbreak} of fibres, suggesting that the
additional fibres arise from splitting
%of another fibre.
a fibre into multiple parts while preserving the total fibre length.

\subsection{Application: Fingerprint data}\label{secfingerprint}

The second application we consider is that of pores lying along ridge
lines in fingerprints.
Fingerprint pore data is considered in some depth in \citet{Tho08} and
\citet{Su09}.

We used a portion of the data set extracted from fingerprint a002--05
from the \textit{NIST} (\textit{National Institute of Standards and Technology})
\textit{Special Database~30} [\citet{Nist}].

%}
%
%fibres and circles indicate points %allocated to noise in this sample]{
%}
%a series of samples of fibres %(darker areas indicate higher
%densities)]{
%}
%
%symbols indicate different clusters, %dots indicate noise. Estimated
%by considering how often pairs of points are associated with the same
%fibre %across a number of samples.]{
%}
%Special Database 30} \citet{Nist}.}

%f6 #&#
%
\begin{figure}
\begin{tabular}{@{}c@{\quad}c@{}}

\includegraphics{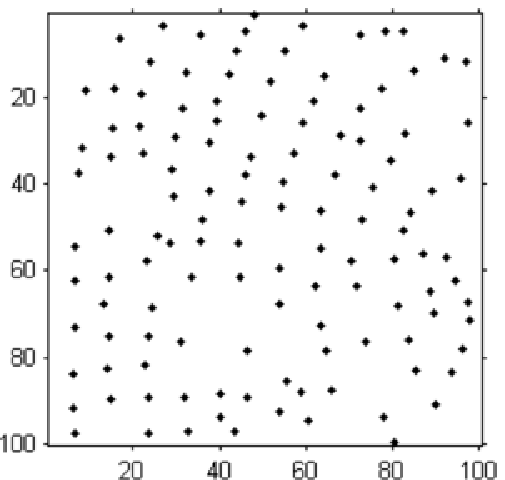}
 & \includegraphics{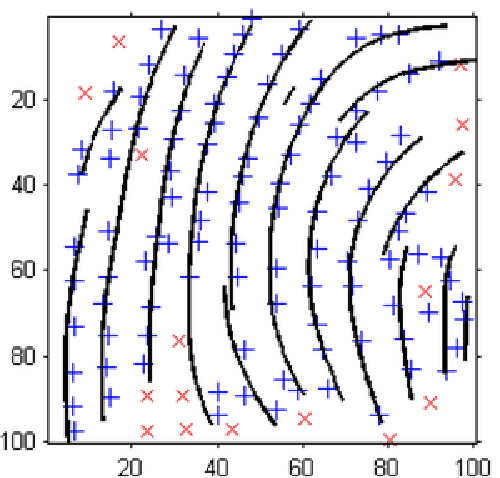}\\
(a) & (b)\\[4pt]

\includegraphics{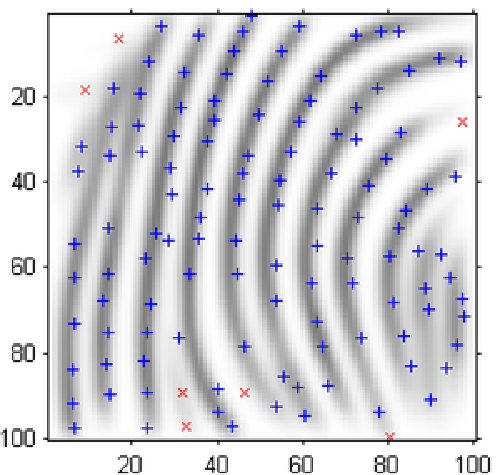}
 & \includegraphics{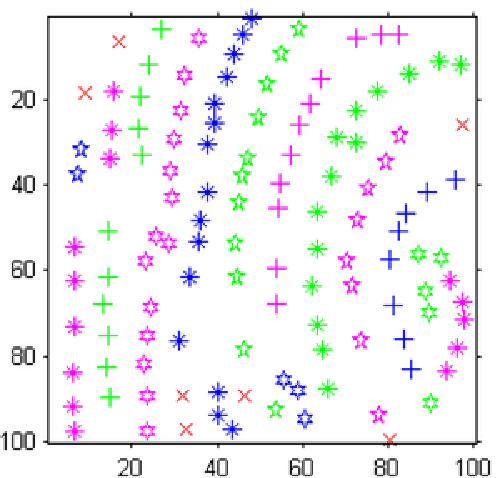}\\
(c) & (d)\vspace*{-3pt}
\end{tabular}
\caption{Pores from portion of fingerprint a002--05 from the \textup{NIST
Special Database 30} [Watson (\protect\citeyear{Nist})]. \textup{(a)} Pore data.
\textup{(b)}
A random sample from the BDMCMC output. Fibres are represented by
curves, pluses indicate points allocated to signal and crosses indicate
points allocated to noise in this sample. \textup{(c)} Estimate of the
density of signal points found by smoothing a series of samples of
fibres (darker areas indicate higher densities). The size of points
representing the data has been reduced to enhance the clarity of the
density estimate. \textup{(d)} Estimate of the clustering of the signal
points---different symbols indicate different clusters, crosses
indicate noise. Estimated by considering how often pairs of points are
associated with the same fibre across a number of samples.}
\label{figeg4}\vspace*{-3pt}
\end{figure}

The birth--death MCMC was run for 40,000 units of algorithm time, the
first 8000 of which were discarded.
Samples were taken at a rate of $0.007$ per unit of time.
The initial state was a randomly sampled set of $\kappa=10$ fibres.

The fingerprint pore data will typically cause breakdown of nearest
neighbor clustering methods.
This is because, while the fibrous structure of the point pattern is
clear when viewing the global picture, it is not so apparent on a small scale.
This phenomena is partly due to the apparent inter-ridge alignment of
points [from left to right in Figure~\ref{figeg4}(a)].
By way of contrast, our field of orientations model takes any
information available on a small scale and uses it across the window,
thanks to the smoothing step in the field of orientations.

As Figure~\ref{figeg4} shows, our model succeeds in fitting many of the
fibres (or fingerprint ridges) to the pore data.
Figure~\ref{figeg4}(c) indicates areas of doubt in the fibre locations
where the shading is lighter near the edges of the window, showing that
fibre samples were more dispersed.

%The curvature bias is barely noticeable in this example (see for
%example the right edge of the window where %fibres are marginally
%straighter than the data suggests), however we have looked at other
%fingerprint data %sets where this bias is more evident.
%Whilst it is important to be aware of this bias it distracts from the
%main ideas so we do not give a visual %example here.

This data set is an ideal candidate for reconstruction of missing data.
We work under the assumptions that pores lie at fairly regularly
intervals along ridges, but some are not identified during the pore
extraction process.
Our method uses information from nearby ridges to complete fibres where
data is missing.
In this example this is particularly evident in the region below the
center of the window.
Knowledge of the posterior distribution of fibres could lead to a
``filling in the gaps'' approach to reconstructing the missing pore
data.\vadjust{\goodbreak}

Table~\ref{tabeg4} gives some numerical properties of the posterior
distribution of fibres.

%data from load ctOut_S_c40530.mat
%39.4 hours
%
%t3 #&#
%
\begin{table}
\caption{Fingerprint Pore Data Set: Posterior means and 50\% and 95\%
credible intervals of~a~selection of properties of the posterior
distribution conditional on the number of~fibres.~The
data was extracted from a portion of fingerprint a002--05 from
the~\textup{NIST}~(\textup{National~Institute of Standards and Technology})
\textup{Special
Database 30} [Watson (\protect\citeyear{Nist})]. It consists of 123 points
on a $100\times
100$ window. A dispersion parameter of $\sigma_{\mathrm{disp}} = 1.5$
is~used, and the mean prior probability a point is
noise~is~$0.091$.~Posterior probabilities only~given if nonzero to rounding error}
\label{tabeg4}
\begin{tabular*}{\tablewidth}{@{\extracolsep{\fill}}
ld{2.2}d{2.2}d{2.2}d{2.2}d{2.2}d{2.2}d{2.2}d{2.2}@{}}
% & posterior mean & 50\% HPD interval & 95\% HPD interval \\
\hline
& \multicolumn{8}{@{}c@{}}{\textbf{Posterior probabilities for number of fibres}}\\
\hline
\multicolumn{1}{@{}l}{Number of fibres}
& 13 & 14 & 15 & 16 & 17 & 18 & 19 & 20 \\
\multicolumn{1}{@{}l}{Posterior probability}
& 0.03 & 0.11 & 0.17 & 0.17 &
0.25 & 0.11 & 0.09 & 0.05 \\
\hline
\end{tabular*}
\begin{tabular*}{\tablewidth}{@{\extracolsep{\fill}}
lcd{3.2}cc@{}}
& \multicolumn{4}{@{}c@{}}{\textbf{Other properties conditioned on the number
of fibres}}\\[-4pt]
& \multicolumn{4}{c@{}}{\hrulefill}\\
& \multicolumn{1}{c}{\textbf{Number of}}
& \multicolumn{1}{c}{\textbf{Posterior}}
& \multicolumn{1}{c}{\textbf{50\% HPD}}
& \multicolumn{1}{c@{}}{\textbf{95\% HPD}}
\\
& \multicolumn{1}{c}{\textbf{fibres} $\bolds{k}$}
& \multicolumn{1}{c}{\textbf{mean}} & \multicolumn{1}{c}{\textbf{interval}}
& \multicolumn{1}{c@{}}{\textbf{interval}}\\
\hline
Number of noise points
& 14 & 14.45 & $[14,16]$ & $[9,18]$ \\
& 15 & 15.71 & $[12,15]$ & $[12,21]$ \\
& 16 & 15.00 & $[12,17]$ & $[8,21]$ \\
& 17 & 18.86 & $[16,19]$ & $[16,23]$ \\
& 18 & 16.08 & $[15,18]$ & $[10,25]$ \\
[4pt]
95th percentile & 14 & 3.76 & $[3.74,3.82]$ & $[3.25,4.22]$ \\
\quad of the distances & 15 & 3.68 & $[3.52,3.69]$ & $[3.32,4.77]$ \\
\quad from signal points & 16 & 3.56 & $[2.34,3.52]$ & $[3.34,3.89]$ \\
\quad to fibres & 17 & 3.58 & $[3.38,3.64]$ & $[3.23,3.95]$ \\
& 18 & 3.67 & $[3.44,3.75]$ & $[3.24,4.38]$ \\
[4pt]
Total length of fibres & 14 & 862.18 & $[861,883]$ & $[785,964]$ \\
& 15 & 882.38 & $[872,933]$ & $[818,966]$ \\
& 16 & 864.75 & $[836,886]$ & $[784,927]$ \\
& 17 & 814.43 & $[788,804]$ & $[788,878]$ \\
& 18 & 861.76 & $[821,876]$ & $[761,941]$ \\
\hline
\end{tabular*}
\end{table}

\section{Discussion}

In this paper we have identified a new model for fibre processes and
for point processes generated from a fibre process.
We have shown how Monte Carlo methods can be used to sample from the
posterior distribution of a fibre process that is instrumental in
generating a point process.

Many different data sets of this type arise in nature.
We investigated earthquakes that cluster around fault lines and pores
in fingerprints that are situated along the fingerprint ridges.
Other data can be found in catalogues of galaxies in the visible universe.
Galaxies are known to align themselves along ``cosmic'' filaments which,
in turn, connect to form a web-like structure.
Understanding these fibres and identifying where they lie is of great
interest to cosmologists; see, for example, \citet{Mar02} for a
statistical overview of some current ideas and also \citet{Sto07} for a
different approach to modeling the filament structure.
Other data sets for which this model may be suitable include the
locations of land mines, often placed in straight lines. Identifying
these lines may aid in the discovery of currently undetected mines.
Similar methods of detecting structure in noisy pictures are a
prominent area of research in image recognition.

This process can be used to fit nonparametric curves to point patterns
with just two limitations on the nature of the curves.
The limitations are that the curves must not intersect, and that they
must be ``sufficiently'' smooth (i.e., there must be no acute angles in
the discretization of the fibres).
The smoothness property is desirable to identify smooth curves rather
than complex structures.
The nonintersection property may be less desirable but, at some
computational cost, the model could be generalized to allow each fibre
to integrate a different field of orientations.

% As we
We
do not make use of a deterministic algorithm (such as the EM-algorithm)
to fit the fibres,
and
our approach is not highly sensitive to the choice of starting parameters.
% and
Therefore, it can be used to provide interval estimates for various parameters.
One of the most sensitive parameters fixed in the algorithm is $\sigma
_{\mathrm{disp}}^2$ which governs the deviation of points from the fibres.
If chosen too large, the result will be too few fibres with a sizeable
error in their locations.
If chosen too small, fibre clusters may be split into multiple parallel
smaller clusters.
Our experience is that the
algorithm is
reasonably
% fairly
robust to changes in other parameters.

One strength of our model is that it fits the noise-signal and cluster
allocations implicitly, in contrast to other cases where the clustering
may need to be predetermined.
The advantage is that we can produce reliability estimates for these
clustering and noise allocations and explore more potential clustering
configurations,
and hence more fibre structures.

A limitation of our model arises from
% are
the constraints on the similarity of fibres.
Fibres are assumed to be of the same width (the displacement of points
from the fibres is independent of the fibre), and have the same mean
points per unit fibre length.
These are not always reasonable assumptions,
% - take for example
as is evidenced by
the earthquake data set.
We could extend the model to allow parameters $\sigma_{\mathrm
{disp}}^2$ and $\eta$ to take different values for each fibre in order
to eliminate this issue.
A further extension would be to include isotropic clusters of points
which do not fit
% under
well to
the ``fibre'' model.

The complexity of the model, considering the infinite dimensionality of
the field of orientations, raises the question of whether or not the
Markov chain adequately explores the sample space.
% In our
Our
examples
indicate that,
% it would seem that,
while the sample space of fields of orientations is not
% adequately
explored
particularly well,
the space of fibre\vadjust{\goodbreak}
configurations
is well explored and the field of orientations varies enough to explore
a wide space of fibre
configurations.
However, as the density of fibres increases, so the MCMC algorithm
requires %substantially
longer runtime in order to overcome these issues.

% Whilst
Note that while
our model performs as well as other available techniques on the basic
data sets,
it demonstrates significantly better performance on the fingerprint
data where a large number of dense fibre clusters account for most of
the data.

It is necessary to bear in mind the ramifications of edge effects in
the model and subsequently the MCMC algorithm.
As we are sampling from a bounded subset $W\subseteq\mathbb{R}^2$, the
omission of potential points just outside $W$ induces a bias on
distance-related measures.
The field of orientations will have a bias at the edge favoring
orientations parallel to the sides of a~rectangular window $W$.
Fibres are created by sampling a random reference point from the field
and integrating the field of orientations from that point.
However, the reference point cannot be sampled from outside $W$, and
fibres that extend past the boundary of $W$ are typically terminated on
the border as no field of orientations is available past that point.
Also, the model for the displacement of points from fibres does not
account for edge effects.
Most of these algorithmic biases would be significantly decreased by
creating a~border around $W$ and completing the analysis over the whole area.
However, this would come at an additional computational cost.

We have commented in passing on the phenomenon of curvature bias and
its effects on the estimation of parameters,
and we note this as
%. Further investigation would be
a fruitful area
% of
for future
research.
Further research possibilities include the fitting of two-dimensional
surfaces in 3 dimensions.
Then new geometric issues need to be taken into account; for example,
it is not the case that a generic field of tangent planes can be
developed into a fibration by surfaces.
It is hoped to investigate this problem in further work.

%suskaldyti doi

% imsref loaded by lrinkeviciute, 2012-05-15 11:43:26
% imsref loaded by lrinkeviciute, 2012-05-15 12:12:18
%

\printaddresses

\end{document}